\documentclass[final,3p,times,twocolumn]{elsarticle}

\usepackage{graphicx}
\usepackage{subfigure} 
\usepackage{amssymb}

\journal{Colloids and Surfaces A}
\begin{document}

\begin{frontmatter}



\title{The characterization of wettability of substrates by liquid nanodrops}


\author{Masao Iwamatsu}
\ead{iwamatsu@ph.ns.tcu.ac.jp}

\address{Department of Physics, Faculty of Liberal Arts and Sciences, Tokyo City University, Setagaya-ku, Tokyo 158-8557, Japan}

\begin{abstract}
Wettability of a substrates is characterized by a contact angle.  Applicability of the simple formula developed by Derjaguin, which relates to the contact angle and disjoining pressure to nano-scale liquid droplets is reconsidered within the framework of the theory of the first-order wetting transition of volatile liquids.  It is concluded that his formula is generally correct for large droplets on an incompletely-wettable substrate.  But it can not be applied to nanodroplets, in particular, on a completely-wettable substrate. An effective contact angle can be defined even for the nanaodroplet.  A formula similar to the Derjaguin  formula is proposed by which we can calculate the contact angle of nanodorplets on both an incompletely- and a completely-wettable substrate for which the whole volume of the droplet is under the influence of the disjoining pressure.
\end{abstract}


\end{frontmatter}


\section{Introduction}
\label{sec:sec1}

\begin{figure}[htbp]
\begin{center}
\includegraphics[width=1.0\linewidth]{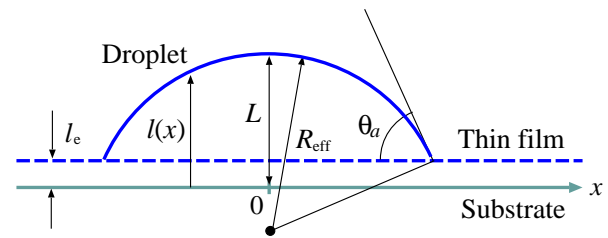}
\caption{
An ideal hemispherical (cylindrical) droplet on a substrate.  The apparent contact angle $\theta_{a}$ is defined as the angle of the intersection of the cylindrical or spherical droplet surface with the effective radius $R_{\rm eff}$ and the height $L$ from the substrata and the wetting layer of the thickness $l_{e}$.
}
\label{fig:1}
\end{center}
\end{figure}

The contact angle, which Young~\cite{Young1805} simply called angle, plays a fundamental role in various fields of surface science.  It has been used to characterize the wetting properties of various surfaces.  Materials deposited on a flat substrate form a hemispherical droplet shown in Fig.~\ref{fig:1} and it is characterized by the radius $R_{\rm eff}$ and the apparent contact angle $\theta_{a}$ by which we can tell that the material can wet or not wet the substrate.  Of course, Fig.~\ref{fig:1} is idealized picture and there is a transition region between the bulk droplet and the thin liquid film.  The information about wettability extracted from the contact angle can then be used to design useful new artificial surfaces and materials. Even though, the contact angle has been used for many years to characterize the surface experimentally or empirically, a reliable theoretical method which can predict the contact angle has still not been well developed.  

Derjaguin~\cite{Derjaguin1987}, more than three decades ago, proposed a formula which relates to the contact angle and the disjoining pressure.  Although, this formula opened a way to predict the contact angle and, therefore, wettability from the information of the disjoining pressure provided theoretically~\cite{Churaev1995,Boinovich2010} or experimentally~\cite{Boinovich2011a,Boinovich2011b}, its utility is not well recognized~\cite{Henderson2005,Evans2009}.
Also, several variants of derivation~\cite{Yeh1999, Starov2007} of the original Derjaguin formula~\cite{Derjaguin1987} have been proposed.  However, due to the recent advances in nan-science and nano-technology, the contact angle of the nano-scale droplet becomes measurable using an atomic force microscope (AFM)~\cite{Checco2008} and environmental scanning electron microscopy (ESEM)~\cite{Mattia2006}.  In contrast, usual optical microscopy can only measure the contact angle of the mili-scale to mico-scale droplet.

In our recent publication~\cite{Iwamatsu2011}, we considered the droplet of a volatile liquid nucleated on a flat substrate which exhibits the first-order wetting transition~\cite{Dietrich1988,Blossey1995}. We pointed out that the droplet is metastable and is in fact the critical nucleus of the heterogeneous nucleation~\cite{Turnbull1950}.  Since the droplet is the critical nucleus of nucleation, it is well know that it can exist only when the surrounding vapor is {\it oversaturated} when the substrate is incompletely wettable and is characterized by the finite macroscopic contact angle.  Therefore, the oversaturated vapor is necessary to maintain the hemispherical meniscus of the droplet. 

On the other hand, it is also well know~\cite{Iwamatsu2011,Starov1979,Starov2004} that the droplet can exist not only in the oversaturated vapor but also in the {\it undersaturated} vapor above the so-called {\it prewetting} line~\cite{Dietrich1988,Blossey1995,Bonn2001}. The droplet in the undersaturated vapor is not the critical nucleus of the bulk vapor-liquid phase transition but that of the surface thin-thick transition~\cite{Dietrich1988,Blossey1995,Bonn2001}.  Therefore, the droplet will not grow infinity but will only grow up to the thickness of the thick wetting film.  The shape of droplet said to deviates significantly from spherical shape and is more appropriately characterized by a "pancake" or "pill box" shape. This drop in the undersaturated vapor has already been considered many years ago~\cite{Starov1979} before the development of the theory of surface phase transition~\cite{Dietrich1988,Bonn2001}, and is called microdrops~\cite{Starov2004}.  It is also possible to define  a microscopic contact angle even though the macroscopic contact angle vanishes as the substrate is completely wettable~\cite{Iwamatsu2011}.  In this article, we will study the contact angle of the droplet which is in fact the critical nucleus of the first-order wetting phase transition.  We will focus on the contact angle for a completely-wettable substrate as well as on an incompletely-wettable substrate.

\section{Morphology of a droplet using the interface-displacement model}
\label{sec:sec2}

\begin{figure}[htbp]
\begin{center}
\includegraphics[width=1.0\linewidth]{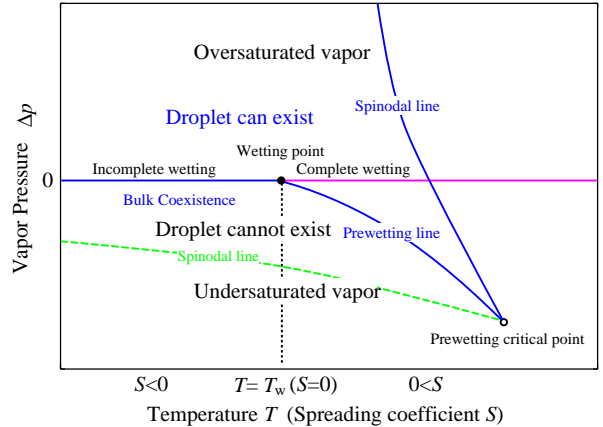}
\caption{
Schematic surface phase diagram in the temperature $T$ and the vapor pressure $\Delta p$ plane.  The droplet can exist only in the oversaturated vapor ($\Delta p>0$) above the bulk coexistence in the incompletely wettable region ($T<T_{W}$).  In the completely wetting region, the droplet can exist even in the undersaturated vapor ($\Delta p<0$) above the prewetting line.  It can also exist in the oversaturated vapor below the spinodal line.  A droplet can exist in the region surrounded by three blue lines. }
\label{fig:2}
\end{center}
\end{figure}

Figure \ref{fig:2} shows the schematic surface phase diagram for the first-order surface phase transition~\cite{Iwamatsu2011,Blossey1995,Bonn2001}.  The horizontal line $\Delta p=0$ corresponds to the bulk coexistence line, on which at the wetting transition point W with the temperature $T=T_{\rm w}$ a first-order wetting transition occurs.  The vapor pressure $\Delta p$ is measured from the pressure at the liquid-vapor coexistence.  The substrate is incompletely wettable with a finite macroscopic contact angle below the wetting transition point W, and is completely wettable with a vanishing macroscopic contact angle above the wetting transition point.  The vapor is oversaturated ($\Delta p>0$) above the coexistence and undersaturated ($\Delta p<0$) below the coexistence.  The prewetting transition line $\Delta p_{\rm p}$ starts at W and terminates at the prewetting critical point.  Above this prewetting line and below the bulk coexistence ($0>\Delta p>\Delta p_{\rm p}$), the stable state is a thick film.  The thickness diverges at the bulk coexistence and this infinitely thick film becomes bulk liquid phase above the coexistence ($\Delta p\ge 0$).  Below the prewetting line ($\Delta p<\Delta p_{\rm p}$), the stable state is a thin film.  

A droplet can exist on the substrate as the critical nucleus only under the oversaturated vapor above the coexistence when the substrate is incompletely wettable (Fig.~\ref{fig:2}).  However, it can exist even in the undersaturated vapor above the prewetting line and below the spinodal line (Fig.~\ref{fig:2}) when the substrate is completely wettable and the macroscopic contact angle vanishes. In this case, the vapor pressure becomes effectively oversaturated due to the disjoining pressure of liquid layer~\cite{Iwamatsu2011}.  Of course, the droplet can also exist in the oversaturated vapor when the substrate is completely wettable as long as the pressure does not exceed the spinodal line.  Therefore, the droplet can exist on the completely-wettable substrate even though the macroscopic contact angle is zero~\cite{Iwamatsu2011}.  Since various properties of the droplet on a completely-wettable substrate have been studied in previous publication~\cite{Iwamatsu2011}, we will pay equal attention to the droplet on an incompletely-wettable substrate.

In order to study the morphology and the contact angle of the droplet on a flat substrate, we will use the interface displacement model (IDM), which has been extensively used to study the wetting transitions~\cite{Yeh1999,Starov2007,Blossey1995,Dobbs1993}.  We will consider only the two-dimensional cylindrical drop which can be handled analytically.  Also, we will not use a gradient-squared approximation which can be used only when the contact angle is small~\cite{Dobbs1993}.  Within the IDM, the free energy functional $\Omega$ of the droplet is written as 
\begin{equation}
\Omega\left[l\right]=\int\left[\gamma \left(\left(1+ \left(\nabla l\right)^{2}\right)^{1/2}-1\right)+V\left( l\right)-\Delta pl\right]dx
\label{Eq:1z}
\end{equation}
where $\gamma$ is the liquid-vapor surface tension, $V\left(l\right)$ is the interface potential~\cite{Yeh1999,Starov2007,Blossey1995,Dobbs1993} from the substrate,  which is related to the so-called disjoining pressure $\Pi\left(x\right)$ through~\cite{Derjaguin1987,Boinovich2011a,Starov2007}
\begin{equation}
V\left(l\right)=\int_{l}^{\infty}\Pi\left(x\right) dx,
\label{Eq:2z}
\end{equation}
and $\Delta p$ denotes the deviation of the vapor pressure from the liquid-vapor coexistence such that $\Delta p=0$ is the bulk liquid-vapor coexistence and for $\Delta p>0$ the vapor is oversaturated. In Eq.~(\ref{Eq:1z}) we keep the nonlinear dependence on $\nabla l$~\cite{Dobbs1993}.  

\begin{figure}[htbp]
\begin{center}
\subfigure[Typical shapes of the full potential $\phi\left(l\right)$.]
{
\includegraphics[width=1.0\linewidth]{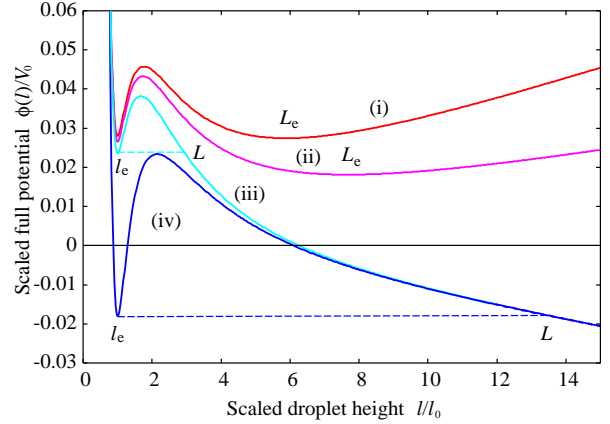}
\label{fig:3a}}
\subfigure[The corresponding disjoining pressure $\Pi\left(l\right)$]
{
\includegraphics[width=1.0\linewidth]{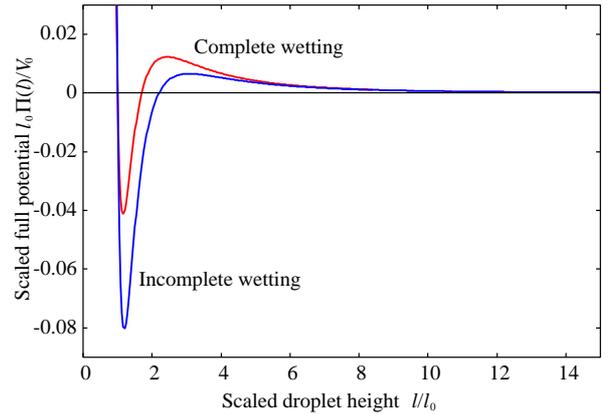}
\label{fig:3b}}
\end{center}
\caption{
(a) Scaled full potential $\phi\left(l\right)/V_{0}$ given by Eq.~(\ref{Eq:21z}) as a function of the scaled film thickness $l/l_{0}$ of the completely-wettable substrate ($b=1.7$)) in the undersaturated vapor ($\Delta p<0$) at the prewetting line ($\Delta pl_{0}/V_{0}=-0.0029$, curve (i)), in the undersaturated vapor above the prewetting line ($\Delta pl_{0}/V_{0}=-0.0015$, curve (ii)), in the oversaturated vapor below the spinodal line ($\Delta pl_{0}/V_{0}=+0.0015$, curve (iii)), and of the incompletely-wettable substrate ($b=2.2$) in the oversaturated vapor ($\Delta pl_{0}/V_{0}=+0.0015$, curve (iv)). Two horizontal lines between the thin-film thickness $l_{e}$ and the height of the droplet $L$ indicates the energy conservation law Eq.~(\ref{Eq:15z}).  (b) Scaled disjoining pressure $\l_{0}\Pi\left(l\right)/V_{0}=-\left(l_{0}/V_{0}\right)dV\left(l\right)/dl$ of the completely-wettable substrate ($b=1.7$) and of the incompletely-wettable substrate ($b=2.2$) that correspond to (a).  
}
\label{fig:3}
\end{figure}

Sometimes it is useful to consider the full potential $\phi(l)$ defined by
\begin{equation}
\phi(l)=V(l)-\Delta pl.
\label{Eq:3z}
\end{equation}
instead of the effective interface potential $V\left(l\right)$ and the pressure contribution $\Delta pl$ separately.  This full potential $\phi(l)$ depends on the pressure $\Delta p$ as well as the temperature $T$.  Figure \ref{fig:3} (a) shows typical shapes of the full potential $\phi(l)$ of four cases: the completely-wettable substrate in the undersaturated vapor ($\Delta p<0$) at the prewetting line (curve (i)), and in the undersaturated vapor above the prewetting line (curve (ii)), and in the oversaturated vapor below the spinodal line (iii), and finally the incompletely-wettable substrate in the oversaturated vapor (curve (iv)). 

The full potential $\phi(l)$ exhibits local minimum at $l_{e}$ that corresponds to a thin wetting films~\cite{Popescu2012}.  Therefore, the droplet is always surrounded by a thin wetting film when there is the first-order wetting transition.  When the vapor is undersaturated ($\Delta p<0$) a double-well shape typical to the first-order thin-thick prewetting transition is observed (Fig.~\ref{fig:3} (a), curves (i) and (ii)).  In this case, two minima at $l_{e}$ and at $L_{e}$ correspond to the metastable thin and the stable thick wetting films. This thick film become infinitely thick ($L_{e}=\infty$) when the vapor is over saturated $\Delta p\ge 0$ (Fig.~\ref{fig:3} (a), curves (iii) and (iv)).

Since the local minimum $V\left(l_{e}\right)$ at $l_{e}$ is related to the spreading coefficient $S$ and the temperature $T$ through~\cite{Blossey1995}
\begin{equation}
\phi\left(l_{e}\right)=S\propto T-T_{\rm w},
\label{Eq:4z}
\end{equation}
where $T_{\rm w}$ is the wetting temperature, and $S$ is defined by
\begin{equation}
S=\gamma_{\rm sv}-\gamma_{\rm sl}-\gamma,
\label{Eq:5z}
\end{equation}
where $\gamma_{\rm sv}$ and $\gamma_{\rm sl}$ are the substrate-vapor and the substrate-liquid surface tensions.  The first-order wetting transition from the incomplete wetting of a thin liquid film with thickness $l_{e}$ ($\phi\left(l_{e}\right)<0=\phi\left(l=\infty\right)$) to the complete wetting of an infinite thickness with $l=\infty$ ($\phi\left(l_{e}\right)>0=\phi\left(l=\infty\right)$) will occur at the wetting transition point W at $T=T_{\rm w}$ in Fig.~\ref{fig:2} (see also Fig.~\ref{fig:3} (a)). Therefore, the complete-wetting regime with $S>0$ is realized above the wetting temperature $T>T_{\rm w}$.  Then, from the Young's formula~\cite{Young1805,Starov2007}
\begin{equation}
\gamma\cos\theta_{a}=\gamma_{\rm sv}-\gamma_{\rm sl}
\label{Eq:6z}
\end{equation}
we have
\begin{equation}
\cos\theta_{a}=1+\frac{S}{\gamma}=1+\frac{\phi\left(l_{e}\right)}{\gamma}.
\label{Eq:7z}
\end{equation}
from Eqs.~(\ref{Eq:4z}) and (\ref{Eq:5z}).  
 
Therefore the apparent contact angle vanishes ($\theta_{a}=0$) in the complete-wetting regime with $S\ge 0$ and the macroscopic droplet cannot exist. Using Eqs.~(\ref{Eq:4z}) and (\ref{Eq:2z}), Eq.~(\ref{Eq:7z}) is written as
\begin{equation}
\cos\theta_{a}=1+\frac{1}{\gamma}\int_{l_{e}}^{\infty}\Pi\left(x\right)dx,
\label{Eq:8z}
\end{equation}
when $\phi\left(l_{e}\right)\simeq V\left(l_{e}\right)$ ($\Delta p\simeq 0$ or $l_{e}\simeq 0$).  This is the well-know Derjaguin formula~\cite{Derjaguin1987}.  The apparent contact angle can be calculable once we know the disjoining pressure $\Pi\left(l\right)$~\cite{Boinovich2010,Boinovich2011a,Boinovich2011b}.   

It should be noted that since we are considering the first-order wetting transition, the monotonically decreasing surface potential $\phi\left(l\right)$~\cite{Dietrich1988} and the disjoining pressure $\Pi\left(l\right)$ of the second order wetting transition will not be considered even though the first order complete wetting of liquid-vapor system is not observed~\cite{Dietrich1988,Bonn2001}. Since the concept of disjoining pressure is so general that its application is not limited to liquid droplets but is extended to the heterogeneous nucleation of solid grains~\cite{Fensin2010,Frolov2011}, our analysis will be useful not only to those who study wetting but to those who are interested in nucleation.

Next, we will briefly touch on the surface phase diagram in Fig.~\ref{fig:2}.  Details will be found in previous publications~\cite{Iwamatsu2011,Blossey1995,Bonn2001}.  In the complete-wetting regime ($S>0, \;\;T>T_{\rm w}$), the prewetting transition $\Delta p_{\rm p}(T)$($<0$) line appears in the undersaturated vapor below the bulk coexistence line $\Delta p=0$ (Fig.~\ref{fig:3}). At the prewetting line $\Delta p=\Delta p_{\rm p}<0$, the full potential $\phi(l)$ has double-minimum shape  (curve (i) in Fig.~\ref{fig:3} (a)), and the two minimums have the same depth:
\begin{equation}
\phi\left(l_{e}\right) = \phi\left(L_{e}\right),
\label{Eq:9z}
\end{equation}
and
\begin{equation}
\left.\frac{d\phi}{dl}\right|_{l_{e}} = \left.\frac{d\phi}{dl}\right|_{L_{e}}=0.
\label{Eq:10z}
\end{equation}
Then the thin (thickness $l_{e}$) and the thick (thickness $L_{e}$) wetting film can coexist at $\Delta p=\Delta p_{\rm p}<0$.  The prewetting line in the undersaturated vapor plays a similar role to the bulk coexistence $\Delta p=0$ in the incomplete-wetting regime ($T<T_{\rm w}$).  When the pressure is increased ($0>\Delta p>\Delta p_{\rm p}$), the thick film with thickness $L_{e}$ becomes stable and the thin film with $l_{e}$ becomes metastable  (curve (ii) in Fig.~\ref{fig:3} (a)).  At and above the bulk coexistence at $\Delta p=0$, the thickness of the stable thick film diverges ($L_{e}\rightarrow\infty$) and it becomes macroscopic bulk liquid phase (curve (iii) in Fig.~\ref{fig:3} (a)). Finally, the metastable thin film loses stability at the upper spinodal in the oversaturated vapor (Fig.~\ref{fig:2}).  

In the incomplete-wetting regime ($S<0, \;\;T<T_{\rm w}$) and in the oversaturated vapor $\Delta p>0$, the bulk liquid phase becomes stable (curve (iv) in Fig.~\ref{fig:3} (a)).  The liquid droplet can always appear as a critical nucleus of the bulk heterogeneous nucleation in the oversaturated vapor.  On the other hand, the thin wetting film with thickness $l_{e}$ becomes stable in the undersaturated vapor $\Delta p<0$.  Therefore, the liquid droplet on the thin film cannot form on an incompletely-wettable substrate in the undersaturated vapor.  

Based on the phase diagram shown in Fig.~\ref{fig:2} and the shape of the potential $\phi(l)$ shown in Fig.~\ref{fig:3}, we can discuss the morphology of the droplet.  For a cylindrical droplet, the Euler-Lagrange equation for the free energy (Eq.~(\ref{Eq:1z})) is simplified to~\cite{Yeh1999,Starov2007,Dobbs1993}
\begin{equation}
\gamma \frac{d}{dx}\left(\frac{l_{x}}{\left(1+l_{x}^{2}\right)^{1/2}}\right)=\frac{dV}{dl}-\Delta p,
\label{Eq:11z}
\end{equation}
where $l_{x}=dl/dx$.  Equation (\ref{Eq:11z}) could be considered as a kind of equation of motion for a classical particle moving in a potential $-\phi(l)$~\cite{Dobbs1993}.

Equation (\ref{Eq:11z}) can be integrated once to give
\begin{equation}
\frac{-\gamma}{\left(1+l_{x}^{2}\right)^{1/2}}=-\gamma\cos\theta\left( l\right)=V\left(l\right)-\Delta pl+C,
\label{Eq:12z}
\end{equation} 
where $C$ is the integration constant and $\cos\theta\left( l\right)$ is the cosine of the angle $\theta(l)$ made between the tangential line of the liquid-vapor surface at the height $l(x)$ and the substrate~\cite{Yeh1999}.  

Near the substrate, the liquid-vapor interface of the droplet will smoothly connect to the surrounding thin liquid film of thickness $l=l_{e}$ with $l_{x}=0$, the integration constant $C$ in Eq.~(\ref{Eq:12z}) will be given by $C=-\gamma-V\left( l_{e}\right)+\Delta pl_{e}$, and the liquid vapor interface will be determined from
\begin{equation}
\frac{-\gamma}{\left(1+l_{x}^{2}\right)^{1/2}}=\left(V\left(l\right)-V\left( l_{e}\right)\right)-\Delta p\left(l-l_{e}\right)-\gamma.
\label{Eq:13z}
\end{equation} 
Similarly, at the top of the droplet with a height $l=L$, we have again $l_{x}=0$ at $l=L$ (Fig.~\ref{fig:2}), and the liquid vapor interface will be determined from an equation similar to Eq.~(\ref{Eq:13z}) with $l_{e}$ replaced by $L$:
\begin{equation}
\frac{-\gamma}{\left(1+l_{x}^{2}\right)^{1/2}}=\left(V\left(l\right)-V\left( L\right)\right)-\Delta p\left(l-L\right)-\gamma.
\label{Eq:14z}
\end{equation}
Since Eqs.~(\ref{Eq:13z}) and (\ref{Eq:14z}) must be identical, we have
\begin{equation}
V\left(l_{e}\right)-\Delta pl_{e} = V\left( L\right)-\Delta pL,\;\;\;\mbox{or}\;\;\;\phi\left(l_{e}\right)=\phi\left(L\right),
\label{Eq:15z}
\end{equation}
which is similar to the energy conservation law for a classical particle whose (pseudo-)equation of motion is given by Eq.~(\ref{Eq:11z}) moving in a potential surface $-\phi(l)$.  Then, the height $L$ of the cylindrical droplet can be determined from Eq.~(\ref{Eq:15z}).  

Therefore the morphology of a cylindrical droplet can be calculated by integrating the equation of motion Eqs.~(\ref{Eq:13z}) or (\ref{Eq:14z}).  The height $L$ of the droplet is determined from Eq.~(\ref{Eq:15z}) (see the horizontal lines in Fig.~\ref{fig:3} (a)).  It is apparent from Fig.~\ref{fig:2} that the droplet can exist not only on an incompletely-wettable substrate but also on a completely-wettable substrate as long as the droplet's height is less than the range of action of surface forces (nanodroplets) even though the macroscopic contact angle vanishes for the latter.  Also, nanodroplets can exist even in an undersaturated vapor below the bulk coexistence and above the prewetting line on a completely wettable substrate.

Using the expansion $V\left(l\right)-V\left( L\right)\simeq (l-L)\left.(dV/dl)\right|_{l=L}$, we have from Eq.~(\ref{Eq:14z})
\begin{equation}
\frac{-\gamma}{\left(1+l_{x}^{2}\right)^{1/2}}=\left(\left.\frac{dV}{dl}\right|_{l=L}-\Delta p\right)\left(l-L\right)-\gamma
\label{Eq:16z}
\end{equation}
near the top of the droplet $l\simeq L$.  If the effective pressure defined by
\begin{equation}
\Delta p_{\rm eff}=-\left.\frac{dV}{dl}\right|_{l=L}+\Delta p=\Pi\left(L\right)+\Delta p
\label{Eq:17z}
\end{equation}
at the top of the droplet with height $L$ is positive, then Eq.~(\ref{Eq:14z}) becomes 
\begin{equation}
\frac{-\gamma}{\left(1+l_{x}^{2}\right)^{1/2}}=-\Delta p_{\rm eff}\left(l-L\right)-\gamma,
\label{Eq:18z}
\end{equation}
whose solution is a semi-circular shape (Fig.~\ref{fig:1})
\begin{equation}
l=\sqrt{R_{\rm eff}^2-x^2}-\left(R_{\rm eff}-L\right)
\label{Eq:19z}
\end{equation}
where $R_{\rm eff}-L$ is the shift of the base of circular interface.  The effective radius $R_{\rm eff}$ is given by the Kelvin-Laplace formula~\cite{Starov2007}
\begin{equation}
R_{\rm eff}=\frac{\gamma}{\Delta p_{\rm eff}}.
\label{Eq:20z}
\end{equation}
Equation (\ref{Eq:20z}) tells us that the circular liquid-vapor surface near the top of the nucleus can be maintained even in the undersaturated vapor with $\Delta p<0$ as long as $\Delta p_{\rm eff}>0$.  This is the reason why a droplet on a completely wettable substrate can exist in the undersaturated vapor.  The effective positive Laplace pressure ($\Delta p_{\rm eff}>0$) results from the repulsive surface potential with positive disjoining pressure ($\Pi\left(L\right)>0$) near the top of the droplet even though the vapor itself is undersaturated. The effective radius $R_{\rm eff}$ diverges at the prewetting line where $\Delta p_{\rm eff}=0$ from Eqs.~(\ref{Eq:17z}) and (\ref{Eq:10z}) since $L=L_{e}$.  Therefore, as noted before, the prewetting line $\Delta p_{\rm eff}=0$ plays the role similar to the coexistence $\Delta p=0$.  Various properties of the droplet on a completely-wettable substrate are found in our previous publication~\cite{Iwamatsu2011}.  

The same argument cannot apply to the incompletely wettable substrates.  A droplet cannot form in the undersaturated vapor with $\Delta p<0$ even if $\Delta p_{\rm eff}> 0$ or $\Pi\left(L\right)>0$ because the energy conservation law Eq.~(\ref{Eq:15z}) cannot be satisfied from Fig.~\ref{fig:2}.  Therefore, on an incompletely-wettable substrate, the effect of the disjoinig pressure is less dramatic

\begin{figure}[htbp]
\begin{center}
\subfigure[The droplet shapes on a completely-wettable substrate.]
{
\includegraphics[width=1.0\linewidth]{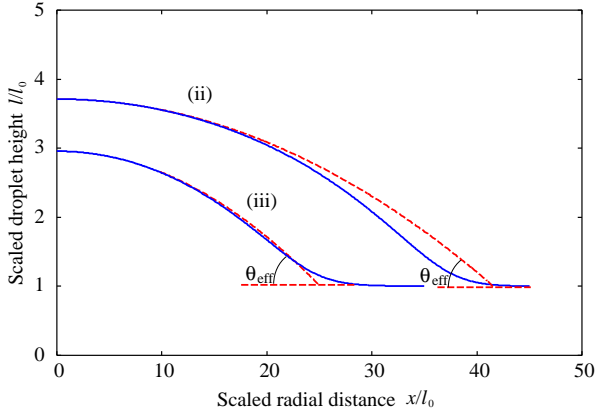}
\label{fig:4a}}
\subfigure[The droplet shapes on a completely-wettable (i) and an incompletely-wettable (iv) substrates.]
{
\includegraphics[width=1.0\linewidth]{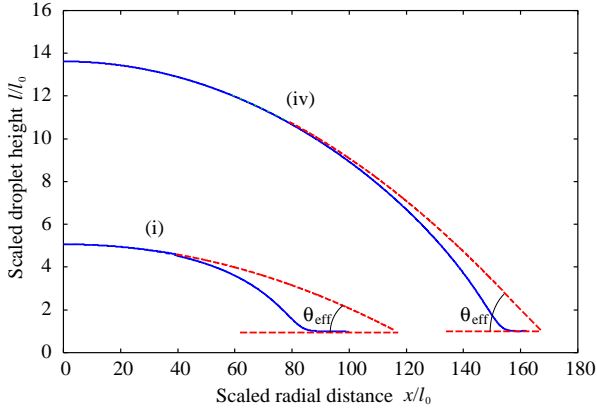}
\label{fig:4b}}
\end{center}
\caption{
The right-half of the droplet shape numerically determined from the Euler-Lagrange equation Eq.~(\ref{Eq:11z}) using the Runge-Kutta method (solid curves) compared with the ideal semi-circular shape given by Eq.~(\ref{Eq:19z}) (broken curves).  (a) Droplet shapes on the completely wettable substrate for the potential (ii) and (iii) of Fig.~\ref{fig:4}.  (b) Those on the completely wettable substrate in the undersaturated vapor at the spinodal and on the incompletely wettable substrate in the oversaturated vapor which correspond to the potentials (i) and (iv) of Fig.~\ref{fig:4}.  The parameter is fixed to $V_{0}/\gamma=0.5$ and other parameters are in the figure caption of Fig.~\ref{fig:3}. 
}
\label{fig:4}
\end{figure}

Figure \ref{fig:4} compares numerically determined cylindrical droplet shapes with the ideal semi-circular shapes (Eq.~(\ref{Eq:19z})) with the height $L$ and the effective radius $R_{\rm eff}$ determined from Eqs.~(\ref{Eq:15z}) and (\ref{Eq:20z}).  The droplet shapes (i) to (iv) in Fig.~\ref{fig:4} (a), (b) corresponds to the full potentials (i) to (iv) in Fig.~\ref{fig:3} (a).  We have used the model potential~\cite{Iwamatsu2011}:
\begin{equation}
V\left(l\right)=V_{0}\left(\frac{1}{2}\left(\frac{l_{0}}{l}\right)^2-\frac{1+b}{3}\left(\frac{l_{0}}{l}\right)^3+\frac{b}{4}\left(\frac{l_{0}}{l}\right)^4\right),
\label{Eq:21z}
\end{equation}
where $l_{0}$ is the thickness of thin wetting layer and $V_{0}$ is related to the Hamaker constant~\cite{Iwamatsu2011,Israelachvili2011}.  Application of more realistic potentials~\cite{Boinovich2010,Boinovich2011a,Henderson2005,Evans2009} would be straightforward. The parameter $b$ controls the complete- and incomplete-wetting of the substrate.  We will have a completely-wettable substrate when $b<2.0$ and an incompletely-wettable substrate when $b>2.0$. By scaling the length by $l_{0}$ and energy by $V_{0}$, we can make all the equations dimensionless.  In Fig.~\ref{fig:3}, we have already shown the potential $\phi\left(l\right)=V\left(l\right)-\Delta pl$ and the corresponding disjoining pressure calculated from Eq.~(\ref{Eq:21z}).  Then, this droplet model is characterized by a single parameter~\cite{Iwamatsu2011} $V_{0}/\gamma$, which we set $V_{0}/\gamma=0.5$ in our calculation.

The droplet is very small on the completely-wettable substrate (Fig.~\ref{fig:4} (a), curve (ii) and (iii)) as it represents microdrops~\cite{Starov2004}.  The droplet shape deviates significantly from an ideal circular shape, in particular, at the prewetting line $\Delta p_{\rm p}$.  The droplet becomes large but flat and its shape becomes pancake-like~\cite{Iwamatsu2011,Bonn2001} as the vapor pressure is decreased down to the prewetting line $\Delta p_{\rm p}$ (Fig.~\ref{fig:4} (b), curve (i)).  The size of the critical pancake is finite even at the prewetting line (Fig.~\ref{fig:4} (b)) even though the effective radius $R_{\rm eff}$ diverges (Eq.~(\ref{Eq:20z})).  On the contrary, the droplet on the incompletely-wettable substrate is large and looks almost spherical in the oversaturated vapor (Fig.~\ref{fig:4} (b), curve (iv)).  This droplet is in fact the critical nucleus of the heterogeneous nucleation~\cite{Iwamatsu2011}.

It is possible to define an effective contact angle $\theta_{\rm eff}$ shown in Fig.~\ref{fig:4} by extrapolating the semi-circular shape (Eq.~(\ref{Eq:19z})) down to the surface of the thin film with thickness $l_{e}$. From the geometrical consideration shown in Fig.~\ref{fig:1}, we find
\begin{equation}
\cos\theta_{\rm eff}=1-\frac{L}{R_{\rm eff}}=1-\frac{\Delta p_{\rm eff}L}{\gamma}
\label{Eq:22z}
\end{equation}
from Eq.~(\ref{Eq:20z}).  This effective contact angle can be defined even on the completely wettable substrate in the undersaturated vapor ($\Delta p<0$) as long as we stay above the prewetting line in Fig.~\ref{fig:2} and $\Delta p_{\rm eff}>0$.  Eq.~(\ref{Eq:22z}) can be transformed further by using Eq.~(\ref{Eq:15z}) as
\begin{equation}
\cos\theta_{\rm eff}=1-\frac{\Delta p_{\rm eff}}{\gamma\Delta p}\left(V\left(l_{e}\right)-V\left(L\right)-\Delta pl_{e}\right),
\label{Eq:23z}
\end{equation}
which can be written using the disjoining potential as
\begin{eqnarray}
\cos\theta_{\rm eff}  =1 &+& \frac{\Delta p_{\rm eff}}{\gamma\Delta p}\left(\int_{l_{e}}^{\infty}\Pi\left(x\right)dx\right. \nonumber \\
&&-\left.\int_{L}^{\infty}\Pi\left(x\right)dx+\Delta pl_{e}\right).
\label{Eq:24z}
\end{eqnarray}
Comparison of Figures. \ref{fig:1} and \ref{fig:4} suggests that the circular droplet shape and the effective contact angle $\theta_{\rm eff}$ calculated from Eq.~(\ref{Eq:22z}) represents the apparent contact angle $\theta_{a}$ in Fig.~\ref{fig:1} fairly well even for the droplet on a completely-wettable substrate except near the prewetting line where the droplet is too flat.

For sufficiently large droplet with sufficiently large $L$ and sufficiently thin wetting film $l_{e}\simeq 0$, one may approximate $V\left(L\right)\simeq 0$ and $\Delta p_{\rm eff}\simeq \Delta p$ from Eq.~(\ref{Eq:17z}).  Then we will recover the original Derjaguin's formula in Eq.~(\ref{Eq:8z}) from Eq.~(\ref{Eq:24z}).  Otherwise, we should use the formula Eq.~(\ref{Eq:24z}), which is the adaptation of Derjaguin's formula for nanodroplets.  The size of the droplet $L$ for which the original Derjaguin's formula is applicable will be around $L\simeq 100$ nm when the dispersion interaction dominates the disjoining pressure since the retardation effect strongly weakens the dispersion interaction for $L>100$ nm~\cite{Israelachvili2011}. Since the droplet on an incompletely-wettable substrate is usually large, we can safely use the original Derjaguin's formula.

Incidentally, the contact angle calculated from the formula Eqs.~(\ref{Eq:22z}) to (\ref{Eq:24z}) are (i) $\theta_{\rm eff}=4.5^{\circ}$, (ii) $\theta_{\rm eff}=12.6^{\circ}$, and (iii) $\theta_{\rm eff}=11.0^{\circ}$ for the droplets on a completely wettable substrate (i) to (iii) in Fig.~\ref{fig:4}.  Of course, the original Derjaguin's formula Eq.~(\ref{Eq:8z}) always predicts $\theta_{\rm eff}=0$ as $S>0$ or $V\left(l_{e}\right)>0$ on a completely wettable substrate.  On the other hand, the effective contact angle for the droplet on an incompletely-wettable substrate (iv) in Fig.~\ref{fig:4} (b) becomes  $\theta_{\rm eff}=9.0^{\circ}$ from the formula Eqs.~(\ref{Eq:22z}) to (\ref{Eq:24z}), while the original Derjaguin's formula predicts $\theta_{a}=7.4^{\circ}$ from Eq.~(\ref{Eq:8z}).  This results suggest that the original Derjaguin's formula is fairly accurate and is a good approximation to the formula Eq.~(\ref{Eq:24z}) for the droplet on an incompletely-wettable substrate though the angles calculated are too small in real experiment which is due to the oversimplification of our model.

We have been considering the cylindrical droplet for which semi-analytical formulas Eqs.~(\ref{Eq:12z})-(\ref{Eq:15z}) can be available since we can integrate the Euler-Lagrange equation once to get Eq.~(\ref{Eq:12z}). Unfortunately, we cannot integrate the Euler-Lagrange equation for an axi-symmetric three dimensional hemispherical droplet~\cite{Iwamatsu2011,Dobbs1999}.  Although we can use Eq.~(\ref{Eq:22z}) for a three dimensional droplet with the effective radius given by~\cite{Iwamatsu2011}
\begin{equation}
R_{\rm eff}=\frac{2\gamma}{\Delta p_{\rm eff}},
\label{Eq:25z}
\end{equation}
for a three dimensional droplet instead of Eq.~(\ref{Eq:20z}), we cannot use the energy conservation law Eq.~(\ref{Eq:15z}) since the Euler-Lagrange equation contains the friction term~\cite{Iwamatsu2011,Blossey1995}.  Therefore, the height $L$ cannot be determined from Eq.~(\ref{Eq:15z}).  Instead, the height $L$ should be determined by solving the Euler-Lagrange equation numerically~\cite{Iwamatsu2011,Dobbs1999}.  Then, Equation~(\ref{Eq:23z}) and, therefore, Eq.~(\ref{Eq:24z}) will have an additional term, and the applicability of the formula Eq.~(\ref{Eq:23z}) as well as the original Derjaguin formula Eq.~(\ref{Eq:8z}) will be more restricted for the three dimensional droplet.


\section{Conclusion}
\label{sec:sec4}

In this paper, we have used the interface displacement model to reconsider the Derjaguin's formula for the contact angle and the disjoining pressure and derived a new formula Eq.~(\ref{Eq:24z}) which can be applicable even for a nanodroplet on a completely-wettable substrate.  On an incompletely wettable substrate in the oversaturated vapor, the droplet is usually large and the Derjaguin's original formula would be applicable.  On the other hand,  on a completely wettable substrate not only in the oversaturated vapor but also in the undersaturated vapor, the droplet can exist as well.  However, the droplet is usually small and highly distorted from sphere.  Even for those small droplets, our new formula Eqs.~(\ref{Eq:23z}) or (\ref{Eq:24z}) can be applicable.  When the droplet is smaller than 100 nm, our new formula is recommendable no matter what the substrate is completely-wetting or  incompletely-wetting since the whole volume of the droplet is under the influence of the disjoining pressure.

\section*{Acknowledgments}
This work was supported by the Grant-in-Aid for Scientific Research [Contract No.(C)22540422] from Japan Society for the Promotion of Science (JSPS) and the MEXT supported program for the Strategic Research Foundation at Private Universities, 2009-2013. 







\end{document}